\documentclass[showkeys,superscriptaddress]{revtex4-2}

\usepackage[]{hyperref}
\usepackage{bbm}
\usepackage{amsfonts}
\usepackage{mathrsfs}
\usepackage{latexsym}
\usepackage{epsfig}
\usepackage{epstopdf}
\usepackage{graphicx}
\usepackage{amssymb}
\usepackage{amsmath}
\usepackage{dcolumn}
\usepackage{bm}
\usepackage{float}
\usepackage{color}
\usepackage{comment}
\usepackage{xcolor}
\begin{document}
\title{   R\'enyi Entropy  and the Topological Features of Charged AdS  Black Hole}
\author{Muhammad Yasir}
\email{yasirciitsahiwal@gmail.com}
\affiliation{Department of Mathematics, Shanghai University  and Newtouch Center for Mathematics of Shanghai University,  Shanghai, 200444, P.R. China}
\author{Tong Lining}
\email{tongln@shu.edu.cn}
\affiliation{Department of Mathematics, Shanghai University  and Newtouch Center for Mathematics of Shanghai University,  Shanghai, 200444, P.R. China}
\author{Farzan Mushtaq}
\email{farzanmushtaq9@gmail.com}
\affiliation{Department of Mathematics, Shanghai University  and Newtouch Center for Mathematics of Shanghai University, Shanghai, 200444, P.R. China}
\author{Xia Tiecheng}
\email{xiatc@shu.edu.cn}
\affiliation{Department of Mathematics, Shanghai University  and Newtouch Center for Mathematics of Shanghai University, Shanghai, 200444, P.R. China}
\author{Sudhaker Upadhyay\footnote{
Visiting Associate, Inter-University Centre for Astronomy and Astrophysics (IUCAA), Pune, Maharashtra 411007,
India}}
\email{sudhakerupadhyay@gmail.com}
\affiliation{Department of Physics, K. L. S. College, Nawada,
Magadh University, Bodh Gaya, Bihar 805110, India}
\affiliation{School of Physics, Damghan University, P.O. Box 3671641167, Damghan, Iran}
\affiliation{Department of General \& Theoretical Physics, L. N. Gumilyov Eurasian National University,  Astana, 010008, Kazakhstan}
\affiliation{Canadian Quantum Research Center 204-3002 32 Ave Vernon, BC V1T 2L7 Canada}

\author{Aram Bahroz Brzo}
\email{arambahroz@gmail.com}
\affiliation{Physics Department, College of Education, University of Sulaimani, Sulaimani 46001, Kurdistan Region, Iraq.}

\author{Behnam Pourhassan}
\email{b.pourhassan@du.ac.ir}
\affiliation{School of Physics, Damghan University, P.O. Box 3671641167, Damghan, Iran.}
\affiliation{Center for Theoretical Physics, Khazar University, 41 Mehseti Street, Baku, AZ1096, Azerbaijan.}
\affiliation{Centre for Research Impact \& Outcome, Chitkara University Institute of Engineering and Technology, Chitkara University, Rajpura, 140401, Punjab, India.}
\affiliation{Physics Department, Istanbul Technical University, Istanbul 34469, Turkey.}

\begin{abstract}{
This study applies Duan's topological current $\phi$-mapping theory to investigate a newly proposed charged modified AdS black hole using the Rényi entropy formalism, examining its thermodynamic properties across the canonical, mixed, and grand canonical ensembles.}
Electric and magnetic charges are fixed in the canonical ensemble, whereas the mixed ensemble incorporates electric and magnetic potentials. The grand canonical ensemble, meanwhile, ensures consistency by employing these potentials exclusively. 
Initially, we compute topological charges by identifying critical points in each ensemble. In both the canonical and mixed ensembles, we identify a conventional critical point with a topological charge of $-1$. We then model the black hole's domain in the AdS regime as a topological defect in thermodynamic space and analyze its local and global topology by calculating the winding numbers at these defect locations. 
{Interestingly, the canonical and mixed ensembles exhibit identical black hole topologies, each possessing a total topological charge of $1$ within the framework of the Rényi entropy.} Finally, we characterize the canonical, mixed, and grand canonical ensembles based on the presence of generation/annihilation points: the canonical ensemble has one such point, the mixed ensemble has a negative one, and the grand canonical ensemble has none.
 \end{abstract}
\keywords{Black hole thermodynamics; Anti-de Sitter (AdS) spacetime; Thermodynamic topology; Modified gravity.}
\maketitle
\section{Introduction}

{The thermodynamic structure of black holes (BHs) in general relativity (GR) and modified theories of gravity is one of the most fascinating and challenging fields of study \cite{1}.} Charged BHs offer an intriguing avenue for understanding the fundamental characteristics of BHs within modified gravity theories. Exploring the thermodynamic topology of charged BHs is particularly interesting, where the interplay between electromagnetic forces and circular motion ensures equilibrium. The thermal properties and behavior of BHs are analyzed using the four well-known laws of BH mechanics \cite{2,3}. The Hawking-Page phase transition \cite{4} establishes a significant connection between thermodynamics and gravity, marking a major discovery in this field. Various approaches have been proposed to study the thermodynamic characteristics of BHs in an extended phase space, where the cosmological constant is interpreted as thermodynamic pressure \cite{5}.
{
Moreover, the Rényi entropy parameter $\lambda$ plays a crucial role in the topology of BH thermodynamics, as it provides a generalized framework for investigating thermodynamic properties, phase transitions, and critical phenomena \cite{6}. Rényi entropy is a generalization of the standard (Boltzmann-Gibbs) entropy.
People use Rényi entropy in black hole thermodynamics to account for non-extensive (non-additive) effects, often inspired by quantum gravity or modified gravity theories.  Rényi entropy has been used in attempts to describe black holes in non-equilibrium thermodynamics or quantum gravitational scenarios.
By varying the Rényi parameter, we examine different aspects of the probability distribution associated with AdS BH microstates, which in turn influence the thermodynamic behavior and topological structure of the system.}

Recently, Wei et al. \cite{7, 8} effectively suggested that BHs might be considered topological thermodynamic defects, expanding on Duan's topological current $\phi$-mapping theory \cite{9}. They correctly categorized various BH solutions according to their global topological charges. The BH solutions can be classified into three distinct topological classes based on their different topological numbers. We can analyze BH solutions and their thermodynamic characteristics from various perspectives. Several studies continue to explore the topological classification of different BHs in this direction.
{The topological numbers are computed for Kerr and Kerr-Newman BHs, and we also discuss the AdS scenario and singly rotating BHs in higher dimensions \cite{10, 11}. Some physicists extend the study of topological classes beyond GR to modified gravity theories such as Lovelock gravity \cite{12} and Gauss-Bonnet gravity \cite{13}. This scenario leads to significant differences in topological numbers, providing new insights into how modified gravity theories and GR differ.}

Cunha, Berti, and Herdeiro studied the stability of light rings (LRs) in ultra-compact objects \cite{14}. They discovered that the LRs of compact objects generally occur in pairs based on the Brouwer degree of a continuous map. This provides a new method for studying LRs by applying a topological argument while neglecting the specific field equations. This finding holds in most cases, with notable exceptions when degenerate light rings appear \cite{15,16,17}.
By calculating the winding numbers of the vector field generated by an effective potential in the orthogonal $(r, \theta)$ space, they established that for each rotation, there exists at least one standard LR outside the  BH horizon \cite{18}.

Moreover, based on their topological argument, an even number of non-degenerate LRs will be visible in a horizonless ultra-compact object, such as a boson star. The observation of non-rotating BHs raises significant concerns regarding the topological argument. This investigation may be extended to BHs in AdS, dS, and asymptotically flat scenarios \cite{14}. Such an analysis can help us understand the topological characteristics of photon spheres (PSs) and reveal the impact of BH spin on the topology of PSs.

These investigations are fascinating because they consistently fall into one of three topological classes, regardless of whether GR, modified gravity, or high-dimensional extensions are considered \cite{19,20}. Furthermore, the number of zero points and the topological number may be significantly influenced by the dimension. Therefore, extending the method to a lower-dimensional scenario might provide a test for universal phenomena or offer an alternative example crucial for further research. A deeper exploration is justified because the topological number of BHs in de Sitter spacetime remains unknown, and the investigation of the topological classification of BHs is still in its early stages. However, a positive cosmological constant provides the most straightforward explanation for this type of accelerated expansion \cite{21}.
{ Notably, unlike the thermodynamics of flat or AdS spacetime, there are two types of horizons in dS spacetime, both of which emit Hawking radiation \cite{22,23}. Thus, studying dS spacetime is essential from both theoretical and practical perspectives \cite{24,25}. Interestingly, the Rényi entropy parameter $\lambda$ can modify the topological structure of these defects, affecting their stability and introducing new critical points. The number and types of critical points in phase space can change, leading to distinct topological classifications as the Rényi parameter $\lambda$ varies.
The study of thermodynamics with Rényi entropy is motivated by its potential to generalize and deepen our understanding of charged AdS black holes beyond the framework of traditional Bekenstein entropy.}

The BHs have been considered topological defects in thermodynamics in \cite{26}. Regarding their topological numbers, BH solutions are classified into three distinct topological classes. {Stationary BHs are regarded as having access to static clarity \cite{27}. Kerr and Kerr-Newman BHs are used to determine the topological classes. Furthermore, the topological numbers of rotating BHs strongly depend on the rotation parameters and spacetime dimensions \cite{27,28}.} Beyond studying topological charges in GR, these investigations are extended to modified gravity theories, including Gauss-Bonnet gravity and Lovelock gravity \cite{29,30,31}. In these cases, the topological charges are entirely different.
It has been demonstrated that free energy is useful for BH topological thermodynamics and for studying BH phase transition dynamics \cite{32,33,34}. The thermodynamic topology of BHs classifies most systems into three categories based on topological numbers: $+1$, $0$, and $-1$. These numbers distinguish between stable and unstable phases of BHs. A well-known relationship expresses thermodynamic pressure $P$ in terms of the cosmological constant, $\Lambda$, as
\begin{equation}\label{i1}
P=-\frac{\Lambda}{8 \pi G},
\end{equation}
where $ G$ denotes Newton's gravitational constant. Consequently, in the extended phase space, the first law of BH thermodynamics takes a modified form, where a thermodynamic volume $V$ is defined as the conjugate variable to the thermodynamic pressure,  $P$
\begin{equation}\label{i2}
d M=T d S+V d P+\sum_i Y_i d x^i,
\end{equation}
where $Y_i d x^i$ represents the $i$-th chemical potential term, $M$ is the total mass, $T$ is the Hawking temperature, and $S$ shows the entropy. A novel approach to understanding the significance of  BHs involves thermodynamic topology. The current $\phi$-mapping (Duan's topological) theory \cite{35,36} is applied within the thermodynamic space of a BH to investigate the effectiveness of this innovative method, which was introduced in \cite{37}. The BH temperature $T$ is expressed as a function of its entropy $S$, state parameter $P$, and other thermodynamic variables as
\begin{equation}\label{i3}
T=T\left(S, P, x^i\right).
\end{equation}
Here, other thermodynamic parameters are represented by \( x^i \). Then, the state parameter is eliminated using the condition 
\[
\left(\frac{\partial T}{\partial S}\right)_{P, x^i} = 0,
\]
leading to the generation of a new potential \( \Phi \), also referred to as Duan's potential
\begin{equation}\label{i4}
\Phi=\frac{1}{\sin \theta} T\left(S, x^i\right).
\end{equation}
A two-dimensional vector $\boldsymbol{\phi} = \left(\phi^S, \phi^\theta\right)$, defined in Duan's $\phi$-mapping theory \cite{35,36}, incorporates a suitability factor of $1/\sin\theta$ and is given by  
\begin{equation}\label{i5}
\phi^S = \left(\partial_S \Phi\right)_{\theta, x^i}, \quad \quad
\phi^\theta = \left(\partial_\theta \Phi\right)_{S, x^i}.
\end{equation}
The zero point of the vector field $\phi$ at $\theta=\pi/2$ is determined by the presence of $\theta$ in $\phi$. These techniques can be used to estimate the critical points. Moreover, the topological charge $j^\mu$ likewise satisfies the conservation law, given by  
\[
\phi^a\left(x^i\right) = 0.
\]  
According to this definition, a topological current exists, and for a given parameter region $\Sigma$, it follows that  
\begin{equation}\label{i6}
Q=\int_{\Sigma} j^0 d^2 x=\sum_{i=1}^N \beta_i n_i=\sum_{i=1}^N w_i,
\end{equation}  
where $j^0$, $w_i$, and $\beta_i$ denote the density of the topological current, the winding number of the $i$-th zero point of $\phi$, and the Hopf index, respectively.
The concept of topology in thermodynamics has been extended to various  BHs \cite{38,39,40,41,42,43} within the framework of \cite{37}. Additionally, the topology of BH thermodynamics has been explored differently in \cite{26}. To begin the analysis, the propagating free energy $F$ is introduced and defined as follows:
\begin{equation}\label{i7}
\mathcal{F} = E - \frac{S}{\tau}.
\end{equation}
Potential energy, entropy, and a quantity with a time dimension are represented by the variables $E$, $S$, and $\tau$, respectively. A vector field $\phi$ in terms of free energy $\mathcal{F}$ follows as 
\begin{equation}\label{i8}
\phi=\left(\frac{\partial \mathcal{F}}{\partial r_{+}},-\cot \Theta \csc \Theta\right).
\end{equation}
The vector $\phi$ has a zero point at $\Theta=\pi/2$.
By adding up the specific winding numbers of each BH charge, one can find the topological number of a BH. The treatment of BHs as topological defects has been extended to include multiple BHs in Refs.~\cite{26, 44,45,46,47,48,49,50,51,52,53,54}.

{
The structure of the paper is organized as follows:
In Sec. \ref{sec2}, we concisely review charged modified BH in the AdS regime and the R\'enyi entropy. 
Sec. \ref{sec4} explores the thermodynamics of BH by analyzing their topological properties within the canonical ensemble.
In Sec. \ref{sec5}, we present a BH solution, interpreted as a topological thermodynamic defect, within the canonical ensemble framework.
Sec. \ref{sec6} delves into the thermodynamic behavior of BHs in the mixed canonical ensemble.
In Sec. \ref{sec7}, we extend the discussion of BH solutions as topological thermodynamic defects in the context of the mixed ensemble.
Sec. \ref{sec8} focuses on studying BHs within the grand canonical ensemble, analyzing their thermodynamic properties.}
In Sec. \ref{sec9}, we examine BH solutions as topological thermodynamic defects within the grand canonical ensemble.
Finally, Sec. \ref{sec10} provides concluding remarks and a summary of key findings.
\section{A Brief Review of charged modified BH in AdS regime and the R\'enyi entropy}\label{sec2}
The topology of charge-modified BH thermodynamics provides valuable insights into the nature of BHs. This approach enhances our understanding of BHs. Furthermore, topology helps in describing phase transitions in a charged BH system. The Einstein-Hilbert action for nonlinear electrodynamics is defined as   \cite{55}
\begin{eqnarray}  \label{R0}
I=\frac{1}{6 \pi} \int_M d^4 x\left(R+\frac{6}{p^2}-4 L\right) \sqrt{-g},
\end{eqnarray}
{where $R$, $L$, $p$, and $g$ represent the Ricci scalar, the Lagrangian associated with a non-electrodynamics theory, the AdS radius, and the determinant of the line element, respectively.}  The Lagrangian $L$ depends on the two invariants listed below,
\begin{eqnarray} \label{R1}
\mathcal{S} =\frac{F_{a b} F^{a b}}{2}, \quad  \mathcal{P}=\frac{F_{a b}(* F)^{a b}}{2},
\end{eqnarray}
where the vector potential $A_a$ and $(* F)_{a b}=\frac{\epsilon_{a b} }{2}F$ are associated with $F_{a b}=\partial_a A_a-\partial_b A_a$. The Einstein-nonlinear electrodynamics equations are
\begin{eqnarray}  \label{R2}
G_{a b}=8 \pi T_{a b}, \quad d * E=0, \quad d F=0.
\end{eqnarray}
Here, $T_{a b}$ and $E_{a b}$ 
  represent the energy-momentum tensor and a nonlinear function of  $\left(* F_{a b}, F_{a b}\right)$, respectively. This can be examined as follows:
\begin{eqnarray} \label{R3}
T_{a b}=\frac{1}{8 \pi}\left(4 F_{a c} F_b^c+2\left(\mathcal{P} L_\mathcal{P}-L\right) g_{a b}\right), \quad E_{a b}=\frac{\partial L}{F_{a b}}=2\left(L_\mathcal{S} F_{a b}+L_\mathcal{P} * F_{a b}\right),
\end{eqnarray}
where $L_\mathcal{S}=\frac{\partial L}{\partial \mathcal{S}}$ and $L_\mathcal{P}=\frac{\partial L}{\partial \mathcal{P}}$, respectively. The Lagrangian for the modified Maxwell theory can be expressed in terms of a dimensionless parameter $\gamma$, which characterizes the modified Maxwell BH solution \cite{56}, as
\begin{eqnarray} \label{R4}
L=\frac{1}{2}\left(\mathcal{S} \cosh \gamma-\sqrt{\mathcal{S}^2+\mathcal{P}^2} \sinh \gamma\right).
\end{eqnarray}
For a non-accelerated BH with modified Maxwell electrodynamics in the AdS regime, the line element is given by
 \cite{56}
\begin{eqnarray} \label{R5}
d s^2=-\frac{f(r) d t^2}{\alpha^2}+\frac{d r^2}{f(r)}+r^2\left(d \theta^2+\sin ^2 \theta \frac{d \phi^2}{K^2}\right),
\end{eqnarray}
where $\alpha$ represents a real constant, and the metric function is given by 
\begin{eqnarray} \label{R6}
f(r)=1-\frac{2 m}{r}+\frac{z^2}{r^2}+\frac{r^2}{p^2},
\end{eqnarray}
here $z^2=e^{-\gamma}\left(q_m^2+q_e^2\right)$. The symbols $m$, $r$, $K$, $q_m$, and $q_e$ denote the mass, radius, conical deficit, magnetic charge, and electric charge of the  {BH}, respectively.
 {The BH solution is singular for $r=p=0$.}
{
The topology of  BH thermodynamics can be investigated using Rényi entropy, as it provides a more generalized framework for studying critical points, phase transitions, and the global structure of the thermodynamic phase space. However, Bekenstein entropy, being a specific case of entropy, does not offer the same flexibility for exploring topological features. The study of Rényi entropy in thermodynamics is motivated by its potential to extend our understanding of charged AdS BHs beyond the traditional Bekenstein entropy. Moreover, Rényi entropy plays a crucial role in the AdS/CFT correspondence, linking BH physics to holographic quantum field theories. Consequently, Rényi entropy serves as a bridge between thermodynamics, geometry, and quantum information, making it essential for understanding the fundamental nature of charged AdS BHs.
}
The area law of the Bekenstein-Hawking entropy \cite{57,58,59} suggests that non-extensive entropy should be used when studying  BH thermodynamics. In non-extensive systems, the BH entropy is referred to as the Tsallis entropy \cite{60}. However, the empirical temperature provided by the Tsallis entropy does not conform to the zeroth law of BH thermodynamics. To resolve this issue, the Tsallis entropy can be modified using the R\'enyi entropy, which can be derived as \cite{6}
\begin{eqnarray} \label{R0}
S_{R}=\frac{1}{\lambda} \ln(1+\lambda S),
\end{eqnarray}
where $\lambda$ is a non-extensive variable with the limit $-\infty < \lambda < 1$. As $\lambda \to 0$, the R\'enyi entropy transforms into the Bekenstein entropy and becomes positive for $0 < \lambda < 1$.

We have reviewed the essential aspects of charged modified BHs in the  AdS regime and introduced the concept of R\'enyi entropy. The non-extensive parameter $\lambda$ in R\'enyi entropy provides a crucial modification to the Bekenstein-Hawking area law, allowing for a more general description of BH thermodynamics. This modification becomes particularly significant when studying phase transitions and critical phenomena. 
The Lagrangian formulation presented here, with its dependence on electric and magnetic invariants, sets the stage for examining the rich topological structure of these BH solutions. The metric function $f(r)$ incorporates both the mass parameter and the effects of the modified Maxwell theory, characterized by the parameter $\gamma$, which will prove essential in our subsequent analysis of thermodynamic topology across different ensembles. This framework provides a robust foundation for investigating the relationship between BH topology and thermodynamic behavior in modified gravity theories.

\section{Thermodynamic Topology and Critical Behavior in the Canonical Ensemble}\label{sec4}
{In the canonical ensemble, electric and magnetic charges are held fixed while examining the thermodynamic properties of a charged modified BH in AdS space.} This ensemble provides a natural starting point for investigating the topological structure of BH thermodynamics, as it maintains constant charges while allowing other parameters to vary. The analysis begins by examining the mass parameter $m$ in the context of Rényi entropy $S_R$, which introduces non-extensive corrections to the standard Bekenstein-Hawking entropy formulation.

The critical behavior in this ensemble arises from the interplay among gravitational attraction, electromagnetic repulsion, and the AdS cosmological constant. This competition manifests in the temperature–horizon radius $(T - r_{h})$ relationship, where multiple phases can coexist under certain conditions. 
{Understanding these phase transitions requires a careful analysis of the topological charges associated with critical points in the thermodynamic parameter space. The study of topological charges in this ensemble is particularly illuminating, as it reveals how the BHs thermodynamic properties are fundamentally connected to the geometric structure of spacetime.
}
The mass parameter in the context of R\'enyi entropy $S_R$ is given by
\begin{eqnarray}\label{1}
m=\frac{\pi  e^{-\gamma } \lambda  p^2 \left(-e^{\gamma }+\pi  \lambda  \left(q_e^2+q_m^2\right)+e^{\gamma +\lambda  {S_R}}\right)+\left(e^{\lambda  {S_R}}-1\right)^2}{2 \pi ^{3/2} p^2 \sqrt{e^{\lambda {S_R}}-1}},
\end{eqnarray}
where $r_h$ represents the horizon radius. The temperature is expressed as a function of pressure, horizon radius, and electric and magnetic charges, as shown below:
\begin{eqnarray}\label{2}
T=\frac{e^{-\gamma } \left(\pi  \lambda  {r_h}^2+1\right) \left(3 e^{\gamma } {r_h}^4-p^2 \left(q_e^2+q_m^2-e^{\gamma } {r_h}^2\right)\right)}{2 \pi  p^2 {r_h}^4}.
\end{eqnarray}
\begin{figure}
\includegraphics[width=19pc]{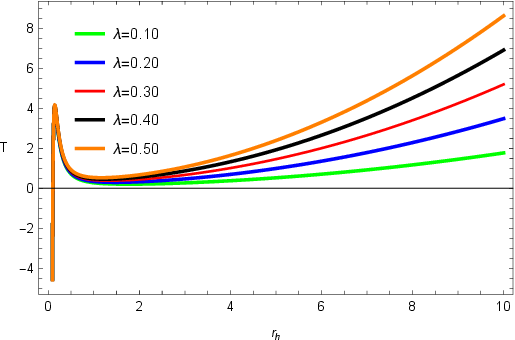}
\caption{\label{f1} Plot of the Hawking temperature $T$  for a BH in canonical ensemble.}
\end{figure}
In Fig. \ref{f1}, we plot the temperature $T$ versus the horizon radius $r_h$ for a charged modified BH in the AdS regime at fixed values of $\gamma = 5.3$, $p = 3$, $q_e = 1$, and $q_m = 1$.
{These curves show that when a first-order phase transition occurs, the exponent $\lambda$ helps in investigating the stability of the BH. It is also noted that the temperature remains positive but initially starts from a negative value, indicating an increasing behavior with the R\'enyi entropy parameter $\lambda$. In this case, we can implement the strategy described in the introduction to examine the parameter space and determine the zero points of the vector field.} As a result, the zero points correspond exactly to the on-shell BH solution. The topological charge can be computed using the topological current $\Phi$-mapping theory. The thermodynamic function $\phi$ can be determined from equation (\ref{i4}) as
\begin{eqnarray}\label{5}
\begin{aligned}
\phi =\frac{e^{-\gamma } \csc (\theta ) \left(\pi  \lambda  {r_h}^2+1\right) \left(3 e^{\gamma } {r_h}^4-p^2 \left(q_e^2+q_m^2-e^{\gamma } {r_h}^2\right)\right)}{2 \pi  p^2 {r_h}^4}.
\end{aligned}
\end{eqnarray}
The vector component of the vector field $\phi=\left(\phi^{r_h}, \phi^\theta\right)$ yields
\begin{eqnarray}\label{6}
\phi^{r_{h}}&=&\left(\frac{\partial \Phi}{\partial r_{h}}\right)_{q_{e, q_m, \theta}}=\frac{3 \lambda  {r_h} \csc (\theta )}{p^2}+\frac{e^{-\gamma } \csc (\theta ) \left(q_e^2 \left(\pi  \lambda  {r_h}^2+2\right)+q_m^2 \left(\pi  \lambda  {r_h}^2+2\right)-e^{\gamma } {r_h}^2\right)}{\pi  {r_h}^5},
\end{eqnarray}
 and
\begin{eqnarray}\label{7}
\phi^\theta=\left(\frac{\partial \Phi}{\partial \theta}\right)_{q_e, q_m, r_{h}}= \frac{e^{-\gamma } \cot (\theta ) \csc (\theta ) \left(\pi  \lambda  {r_h}^2+1\right) \left(p^2 \left(q_e^2+q_m^2-e^{\gamma } {r_h}^2\right)-3 e^{\gamma } {r_h}^4\right)}{2 \pi  p^2 {r_h}^4}.
\end{eqnarray}
 The normalized vector components are
\begin{eqnarray}\label{8}
\frac{\phi^{r_h}}{\|\phi\|}=\frac{2A}{\sqrt{4 A^2+{r_h^2 e^{-2 \gamma } \cot ^2(\theta )  \left(\pi  \lambda  {r_h}^2+1\right)^2 \left(B-3 e^{\gamma } {r_h}^4\right)^2}}},
\end{eqnarray}
  and
\begin{eqnarray}\label{9}
\frac{\phi^\theta}{\|\phi\|}=\frac{ r e^{-\gamma } \cot (\theta )  \left(\pi  \lambda  {r_h}^2+1\right) \left(B-3 e^{\gamma } {r_h}^4\right)}
{\sqrt{4 A^2+{r_h^2 e^{-2 \gamma } \cot ^2(\theta )  \left(\pi  \lambda  {r_h}^2+1\right)^2 \left(B-3 e^{\gamma } {r_h}^4\right)^2}}},
\end{eqnarray}
where
\begin{eqnarray}\label{8A}
 A= 3 \lambda \pi  {r_h}^6+  {p^2} e^{-\gamma }  \left(q_e^2 \left(\pi  \lambda  {r_h}^2+2\right)+q_m^2 \left(\pi  \lambda  {r_h}^2+2\right)-e^{\gamma } {r_h}^2\right),
\end{eqnarray}
\begin{eqnarray}\label{8B}
B= p^2 \left(q_e^2+q_m^2-e^{\gamma } {r_h}^2\right).
\end{eqnarray}
A normalized vector $n=\left(\frac{\phi^{r_h}}{\|\phi\|}, \frac{\phi^\theta}{\|\phi\|}\right)$ appeared in Eq. (\ref{8}) and (\ref{9}).
\begin{figure}
\includegraphics[width=18pc]{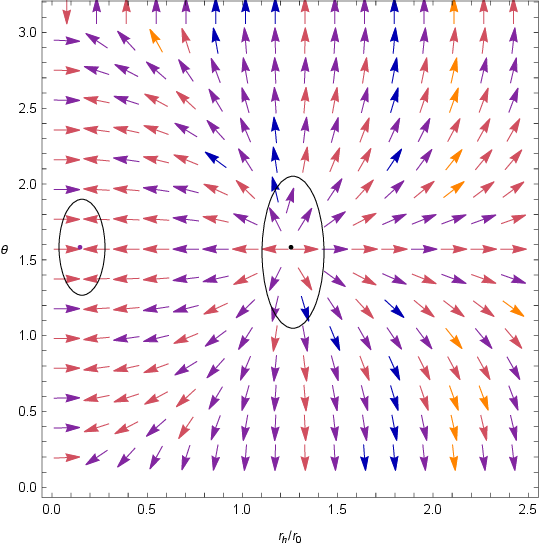}
\caption{\label{f2} Plot of normalized vector field $n$ in $r_h/r_0$ vs $\theta$ plane for BH in canonical ensemble with fixed values  of $\lambda =0.3$, $\gamma =5.3$,  $p=3$, $q_e=1$ and $q_m=1$. The dot represents the critical point.}
\end{figure}
 In {Fig.~\ref{f2}}, we plot the normalized vector in the plane $r_{h}$ versus $\theta$ for a  BH in a charged modified BH in the AdS regime. These unit vectors can be used to locate zero points. Here, we fix the values of $\gamma = 5.3$ and $p = 3$. In this case, $r_{0}$ represents an arbitrary length scale based on the BH size surrounding the cavity. 
{The dot indicates the critical points, defined as $CP_{1}$ (incoming flow) and $CP_{2}$ (outgoing flow).} To compute the critical points, we set $\theta = \frac{\pi}{2}$ in Eq.~(\ref{8}).
\begin{figure}
\includegraphics[width=18pc]{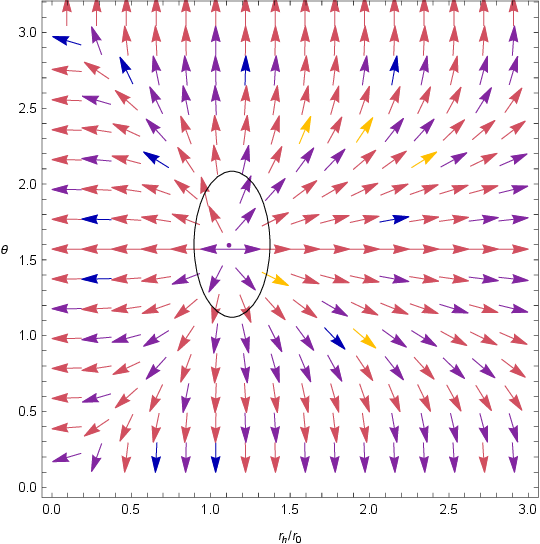}
\caption{\label{f3} Plot of unit vector field $n$ in $r_h/r_0$ vs $\theta$ plane for BH thermodynamic canonical ensemble with  $\gamma =0.3$,  $\lambda =0.5$,  $p=3$,  $q_e=1$, $q_m=1$ and $\tau =30$. The dot represents the critical point.}
\end{figure}
{In {Fig.~\ref{f3}}, we depict the normalized vectors in the plane $r_{h}$ versus $\theta$ by fixing the values $\gamma = 0.3$, $\lambda = 0.5$, and $p = 3$. The dot indicates the critical point, denoted as $CP_{3}$, which represents the outgoing flow leading to an unstable phase.} To calculate the critical points, we set $\theta = \frac{\pi}{2}$ in (\ref{8}).
\begin{figure}
\includegraphics[width=19pc]{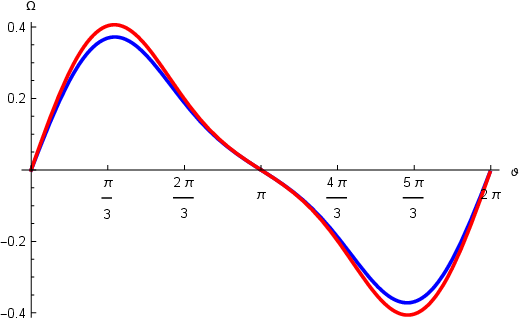}
\caption{\label{f3a} Plot of $\Omega$ vs $\vartheta$ for the contours $C_1$ and $C_2$ of fixed values  $\gamma =1.3$,  $\lambda =0.1$,  $p=2.2$, $q_e=1$, $q_m=1$, $\tau =30$, $a=0.15$ and $b=0.12$.}
\end{figure}
{ One can obtain the topological charge of the critical points by considering a contour $C$ parameterized by $\vartheta \in (0, 2\pi)$, which is defined as follows \cite{61}}
\begin{eqnarray}\nonumber
r_{+}= a cos\vartheta +r_{0},
\theta= b sin\vartheta +\frac{\pi}{2}.
\end{eqnarray}
In {Fig.~\ref{f3a}}, we plot two contours, $C_{1}$ and $C_{2}$, for fixed values of $\gamma = 1.3$, $\lambda = 0.1$, $p = 2.2$, $q_e = 1$, $q_m = 1$, $\tau = 30$, $a = 0.15$, and $b = 0.12$. For these contours, we choose $(a, b, r_{0}) = (0.6, 0.2, 2\sqrt{3})$ and $(0.6, 0.4, 5)$. Here, the red and blue curves represent the contours $C_{1}$ and $C_{2}$, respectively. 

Across the contour $C$, the deflection of the vector field $n$ is given by
\begin{eqnarray}\nonumber
 \Omega (\vartheta) =\int^{\vartheta}_{0} \epsilon_{ab} n^{a}\partial_{\vartheta} n^{b} d\vartheta.
\end{eqnarray}
The topological charge is given by 
\[
Q = \frac{1}{2\pi} \Omega (2\pi).
\] 
The topological charge enclosed by the contours \( C_1 \) and \( C_2 \) is estimated to be zero for both cases. These correspond to the conventional critical points. Thus, the total topological charge is 
$Q = 0$ within both contours. The function \( \Omega (\vartheta) \) for \( C_1 \) and \( C_2 \) approaches zero as \( \vartheta \to 2\pi \).

\section{BH solution as topological thermodynamic defects in canonical ensemble}\label{sec5}

We are currently examining the BH solution in the canonical ensemble as a topological thermodynamic defect. The generalized free energy can be calculated using Eqs.~(\ref{R0}), (\ref{1}), and (\ref{i7}) as
\begin{eqnarray}\label{12}
F= \frac{\frac{{r_h}^4}{p^2}+e^{-\gamma } \left(q_e^2+q_m^2\right)+{r_h}^2}{2 {r_h}}-\frac{\log \left(\pi  \lambda  {r_h}^2+1\right)}{\lambda  \tau}.
\end{eqnarray}
The components of the vector field, as defined by (\ref{i8}), are
\begin{eqnarray}
\phi^{r_h}&=& \frac{1}{2} \left(\frac{3 {r_h}^2}{p^2}-\frac{e^{-\gamma } \left(q_e^2+q_m^2\right)}{{r_h}^2}-\frac{4 \pi  {r_h}}{\pi  \lambda  {r_h}^2 \tau +\tau }+1\right),\label{13}\\
\phi^{\theta} &=&-\cot (\theta) \csc (\theta).\label{14}
\end{eqnarray}
The unit vectors corresponding to the above components of the vector field are
\begin{eqnarray}\label{15}
n^1=\frac{Y-\frac{e^{-\gamma } \left(q_e^2+q_m^2\right)}{{r_h}^2}+1}{2 \sqrt{\cot ^2(\theta ) \csc ^2(\theta )+\frac{1}{4}
\left(Y-\frac{e^{-\gamma } \left(q_e^2+q_m^2\right)}{{r_h}^2}+1\right)^2}},
\end{eqnarray}
and
\begin{eqnarray}\label{16}
n^2=\frac{\cot (\theta ) \csc (\theta )}{\sqrt{\cot ^2(\theta ) \csc ^2(\theta )+\frac{1}{4}
\left(Y-\frac{e^{-\gamma } \left(q_e^2+q_m^2\right)}{{r_h}^2}+1\right)^2}}.
\end{eqnarray}
To obtain the analytical expression for $\tau$, set $\phi^{r_h} = 0$. The corresponding zero points are given as follows:
\begin{eqnarray}\label{17}
\tau=\frac{4 \pi  e^{\gamma } p^2 {r_h}^3}{\left(\pi  \lambda  {r_h}^2+1\right) \left(3 e^{\gamma } {r_h}^4-p^2 \left(q_e^2+q_m^2-e^{\gamma } {r_h}^2\right)\right)}.
\end{eqnarray}
\begin{figure}
\begin{minipage}{18pc}
\includegraphics[width=18pc]{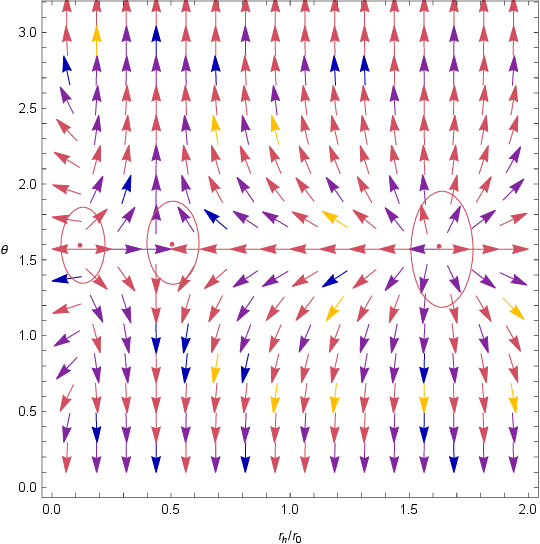}
\caption{\label{f4} Plot of unit vector field $n$ in $r_h/r_0$ vs $\theta$ plane for BH thermodynamic defects in canonical ensemble with  $\gamma =5.3$,  $\lambda =0.1$, $p=2.5$, $q_e=1$, $q_m=1$, $\tau =5$ and $a=0.1$. The dot represents the critical point.}
\end{minipage}\hspace{3pc}%
\begin{minipage}{18pc}
\includegraphics[width=18pc]{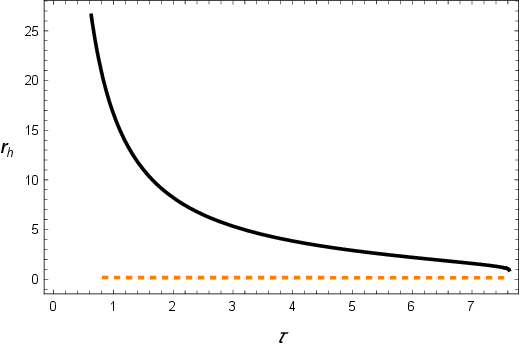}
\caption{\label{f5} { Plot of  $r_h/r_0$ vs $\theta$ plane for BH thermodynamic defects in canonical ensemble with $\lambda =0.5$,  $\gamma =1.3$, $p=0.1$, $q_e=1$ and $q_m=1$.}}
\end{minipage}\hspace{3pc}%
\end{figure}
In the same way, we find the critical points and set $\theta = \frac{\pi}{2}$ in {\ref{15}}.  
In {Fig. \ref{f4}}, we show the $n$ versus $\theta$ plane for the BH thermodynamic defects in the canonical ensemble with $\gamma = 5.3$, $\lambda = 0.1$, $p = 2.5$, $q_e = 1$, $q_m = 1$, $\tau = 5$, and $a = 0.1$. Here, we observe three critical points, denoted as $CP_{4}$, $CP_{5}$, and $CP_{6}$, occurring at $(a, b, r_{0}) = (0.26, 0.18, 0.6)$, $(0.26, 0.18, 0.8)$, and $(0.26, 0.18, 1.7)$, respectively.
In {Fig. \ref{f5}}, we plot $r_h/r_0$ versus $\theta$ for fixed values of $\lambda = 0.5$, $\gamma = 1.3$,  $p = 0.1$, $q_e = 1$, and $q_m = 1$.
 {The size of the cavity surrounding the  BH determines the arbitrary length scale $r_{0}$, and the pressure remains below the critical pressure $P_{c}$. We have studied the pressure values below the critical pressure $P_c$, maintaining the same charge and pressure configuration as in $\tau$. The stability of the AdS BH is investigated, as it is straightforward to demonstrate that a BH with a large radius is stable (having positive free energy), whereas one with a small radius is unstable (having negative free energy).
}
\begin{figure}
\includegraphics[width=19pc]{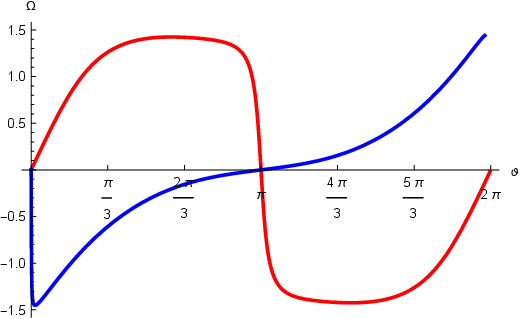}
\caption{\label{f5a} Plot of $\Omega$ vs $\vartheta$ for the contours $C_3$ and $C_4$ with fixed values of  $\gamma =0.3$,  $\lambda =0.5$,  $p=3$, $q_e=1$, $q_m=1$, $\tau =30$, $a=0.26$ and $b=0.18$.}
\end{figure}
{In {Fig. \ref{f5a}}, we construct two contours $C_{3}$ and $C_{4}$ for fixed values of $\gamma =0.3$,  $\lambda =0.5$,  $p=3$, $q_e=1$, $q_m=1$, $\tau =30$, $a=0.26$ and $b=0.18$. Here, red and blue curves are defined by $C_{3}$ and $C_{4}$ contours, respectively. The contour $C_{4}$ has a topological charge of $-1$, indicating the conventional critical point. On the other hand, contour $C_{3}$ lacks a critical point leading to the zero topological charge. Hence, the overall topological charge is $-1$.}

\section{BH in mixed ensemble  }\label{sec6}

In this case, the electric potential $\phi_e$ and the magnetic charge $q_m$ remain constant. The electric potential $\phi_{e}$ is expressed as
\begin{eqnarray}\label{18}
\phi_e=\frac{q_e}{r_h}.
\end{eqnarray}
The mass parameter is then described as
{
\begin{eqnarray}\label{19}
m=\frac{{r_h^4}/{p^2}+e^{-\gamma } \left(q_m^2+r_h^2 {\phi_e}^2\right)+r_h^2}{2 r_h}.
\end{eqnarray}}
We claim that the conventional critical point associated with a topological charge of $-1$ can induce a first-order phase transition. However, the emergence of a new critical point associated with a topological charge of $+1$ does not necessarily indicate the occurrence of a nearby first-order phase transition. The modified temperature can be expressed as
\begin{eqnarray}\label{20}
T=\frac{e^{-\gamma } \left(\pi  \lambda  {r_h}^2+1\right) \left(3 e^{\gamma } {r_h}^4-p^2 \left(q_m^2-{r_h}^2 \left(e^{\gamma }
-{\phi_e}^2\right)\right)\right)}{2 \pi  p^2 {r_h}^4}.
\end{eqnarray}
The temperature expression (\ref{20}) encapsulates the interplay between the modified Maxwell theory, represented by the parameter $\gamma$, and the Rényi entropy corrections, governed by $\lambda$, in the AdS BH thermodynamics. The thermodynamic function $\phi$ is given as
\begin{eqnarray}\label{21}
\phi =\frac{T}{\sin (\theta)}=\frac{e^{-\gamma } \left(\pi  \lambda  {r_h}^2+1\right) \left(3 e^{\gamma } {r_h}^4-p^2 \left(q_m^2-{r_h}^2 \left(e^{\gamma }
-{\phi_e}^2\right)\right)\right)}{2 \pi  p^2 {r_h}^4 \sin (\theta )}.
\end{eqnarray}
The components of the vector field $\phi = \left(\phi^{r_h}, \phi^\theta\right)$ are
\begin{eqnarray}
\phi^{r_h}&=&\frac{3 \lambda  {r_h} \csc (\theta )}{p^2}+\frac{e^{-\gamma } \csc (\theta ) \left(q_m^2 \left(\pi  \lambda  {r_h}^2+2\right)-{r_h}^2 \left(e^{\gamma }-{\phi_e}^2\right)\right)}{\pi  {r_h}^5},\label{22}\\
\phi^{\theta}&=&-\frac{e^{-\gamma } \cot (\theta ) \csc (\theta ) \left(\pi  \lambda  {r_h}^2+1\right) \left(3 e^{\gamma } {r_h}^4-p^2 \left(q_m^2-{r_h}^2 \left(e^{\gamma }-{\phi_e}^2\right)\right)\right)}{2 \pi  p^2 {r_h}^4}.\label{23}
\end{eqnarray}
{To analyze the topological behavior of the BH system, it is necessary to normalize the vector field components. By computing the magnitude of the vector field $\phi$ and dividing each component by this magnitude, one obtains the normalized components, which preserve directional information while maintaining unit length. The radial components of these normalized vector fields are given by}
\begin{eqnarray}
\frac{\phi^{r_h}}{\|\phi\|}&=& \frac{2 \left(3 \lambda \pi {r_h}^5+p^2 e^{-\gamma } \left(q_m^2 \left(\pi  \lambda  {r_h}^2+2\right)-{r_h}^2 \left(e^{\gamma }-{\phi_e}^2\right)\right)\right)}{\sqrt{C+e^{-2 \gamma } {r_h}^2 \cot ^2(\theta ) \left(\pi  \lambda  {r_h}^2+1\right)^2 \left(p^2 \left(q_m^2-{r_h}^2 \left(e^{\gamma }-{\phi_e}^2\right)\right)-3 e^{\gamma } {r_h}^4\right)^2}},\label{24}\\
\frac{\phi^\theta}{\|\phi\|}&=& -\frac{e^{-\gamma } {r_h} \cot (\theta ) \left(\pi  \lambda  {r_h}^2+1\right) \left(3 e^{\gamma } {r_h}^4-p^2 \left(q_m^2-{r_h}^2 \left(e^{\gamma }-{\phi_e}^2\right)\right)\right)}{\sqrt{C+e^{-2 \gamma } {r_h}^2 \cot ^2(\theta ) \left(\pi  \lambda  {r_h}^2+1\right)^2 \left(p^2 \left(q_m^2-{r_h}^2 \left(e^{\gamma }-{\phi_e}^2\right)\right)-3 e^{\gamma } {r_h}^4\right)^2}}.\label{25}
\end{eqnarray}
where $C$ is defined in the Appendix.
 {These normalized components play a crucial role in identifying the critical points and determining the topological charges of the system. Normalization ensures that the magnitude of the vector field remains uniform while preserving the essential topological information encoded in its direction. The complexity of this expression reflects the interplay between gravitational effects and electromagnetic charges. Moreover, modifications are introduced in both the modified Maxwell theory and Rényi entropy.}
\begin{figure}
\begin{minipage}{18pc}
\includegraphics[width=18pc]{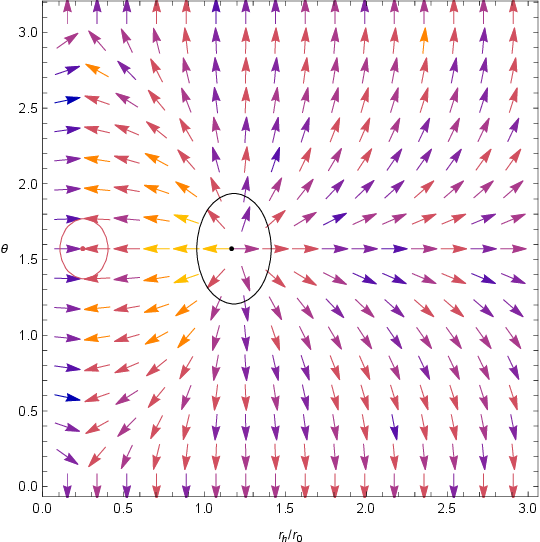}
\caption{\label{f6} Plot of unit vector field $n$ in $r_h/r_0$ vs $\theta$ plane for BH thermodynamic in mixed ensemble
with fixed values of  $\gamma =0.13$,  $\lambda =0.5$,  $p=3.2$, $q_e=1$, $q_m=1$ and $\tau =30$. The dot represents the critical point.}
\end{minipage}\hspace{3pc}%
\begin{minipage}{18pc}
\includegraphics[width=18pc]{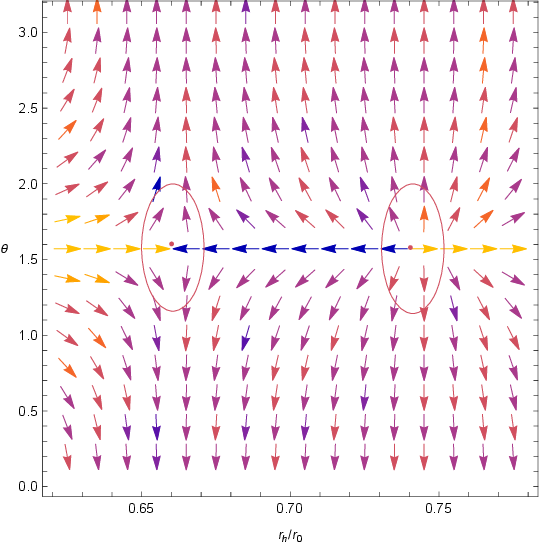}
\caption{\label{f7} Plot of unit vector field $n$ in $r_h/r_0$ vs $\theta$ plane for BH thermodynamic defects in mixed ensemble with fixed values of  $\gamma =0.13$,  $\lambda =0.9$,  $p=3.2$, $q_e=1$, $q_m=1$, $\tau =30$, $a=0.13$ and $b=0.78$.}
\end{minipage}\hspace{3pc}%
\end{figure}
In {Fig.~\ref{f6}}, we represent the unit vector field $n$ for an arbitrarily chosen value of $\tau$, specifically $\tau = 30$, with $\gamma = 0.13$, $\lambda = 0.5$, and $p = 3.2$. We show that there is one critical point, $CP_7$, located at $(a, b, r_0) = (0.13, 0.78, 1.2)$.  
In {Fig.~\ref{f7}}, we plot the unit vector field in the $\theta$ plane, taking $\gamma = 0.13$, $\lambda = 0.9$, and $p = 3.2$. In this case, for $\tau = 30$, we find two critical points, denoted as $CP_8$ and $CP_9$, located at $(a, b, r_0) = (0.13, 0.78, 0.68)$ and $(0.13, 0.78, 0.74)$, respectively.

\begin{figure}
\includegraphics[width=19pc]{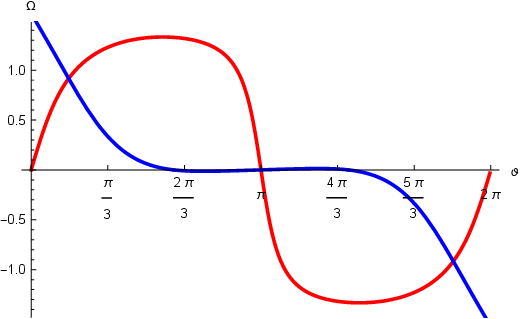}
\caption{\label{f7a} Plot of $\Omega$ vs $\vartheta$ for the contours $C_5$ and $C_6$ with fixed values of  $\gamma =2.3$, $\lambda =0.5$,  $p=3.2$, $q_e=1$, $q_m=1$, $\tau =30$, $a=0.13$ and $b=0.78$.}
\end{figure}
In {Fig. \ref{f7a}}, we have two contours, $C_{5}$ and $C_{6}$, representing the red and blue curves, respectively, with fixed values $\gamma = 2.3$, $\lambda = 0.5$, $p = 3.2$, $q_e = 1$, $q_m = 1$, $\tau = 30$, $a = 0.13$, and $b = 0.78$. 

The contour $C_{5}$ initially increases for certain values and then decreases, intersecting the $\theta$-axis at $\pi$. After forming a loop, $\Omega$ vanishes, implying that the charge is zero. The topological charge along contour $C_{6}$ is $-1$, indicating a conventional critical point.

The pressure is calculated below the critical pressure $P_c$, maintaining the same charge and pressure configuration as in the previous case. The essential points are displayed in the phase structure ($P < P_c$), represented as dots. In this scenario, the small and large  BH  phases are separated by stable and unstable regions (slope sections of the phase structure), with two extremal points corresponding to each isobaric curve.

\section{Topological thermodynamic defects in mixed ensemble}\label{sec7}

This section investigates the modified charged AdS  BH as a topological defect in the mixed ensemble. Nevertheless, we begin with the generalized free energy potential as follows:
\begin{eqnarray}\label{27}
F=m-q_{e} {\phi_e}-\frac{S}{\tau}.
\end{eqnarray}
In this case, the modified mass parameter and temperature can be expressed as follows:
\begin{eqnarray}\label{28}
m=\frac{\frac{{r_h}^4}{p^2}+e^{-\gamma } \left(q_m^2+{r_h}^2 {\phi_e}^2\right)+{r_h}^2}{2 {r_h}},
\end{eqnarray}
and
\begin{eqnarray}\label{28a}
T=\frac{e^{-\gamma } \left(\pi  \lambda  {r_h}^2+1\right) \left(3 e^{\gamma } {r_h}^4-p^2 \left(q_m^2-{r_h}^2 \left(e^{\gamma }-{\phi_e}^2\right)\right)\right)}{2 \pi  p^2 {r_h}^4}.
\end{eqnarray}
With the help of relation~(\ref{i8}), the generalized free energy potential can be obtained as
\begin{eqnarray}\label{29}
F= \frac{\frac{{r_h}^4}{p^2}+e^{-\gamma } \left(q_m^2+{r_h}^2 {\phi_e}^2\right)+{r_h}^2}{2 {r_h}}-\frac{\log \left(\pi  \lambda  {r_h}^2+1\right)}{\lambda  \tau }-{r_h} {\phi_e}^2.
\end{eqnarray}
This can be stated in terms of potentials as
\begin{eqnarray}\label{30}
\phi^{{r_h}}&=& \frac{1}{2} \left(e^{-\gamma } \left({\phi_e}^2-\frac{q_m^2}{{r_h}^2}\right)+Y-2 {\phi_e}^2+1\right).,\\
\label{31}
\phi^{\theta}&=&-\cot (\theta ) \csc (\theta ),
\end{eqnarray}
where $Y$ is defined in the Appendix. 
\begin{eqnarray}
\end{eqnarray}
In this scenario, the intersection points coincide for larger values of $\tau$. The critical points associated with the annihilation point are easily identified. If the winding numbers of the two zero points are given by $\omega_1 = -1$ and $\omega_2 = 1$, then the global topological number for the BH under the modified charged AdS BH is  
$\omega = \omega_1 + \omega_2 = 0.  $

The appropriate unit vectors are
\begin{eqnarray}\label{32}
n^{1}= \frac{Y+e^{-\gamma } \left({\phi_e}^2-\frac{q_m^2}{{r_h}^2}\right)-2 {\phi_e}^2+1}{2 \sqrt{\cot ^2(\theta ) \csc ^2(\theta )+\frac{1}{4} \left(Y+e^{-\gamma } \left({\phi_e}^2-\frac{q_m^2}{{r_h}^2}\right)-2 {\phi_e}^2+1\right)^2}},
\end{eqnarray}
and
\begin{eqnarray}\label{33}
n^{2}= -\frac{\cot (\theta ) \csc (\theta )}{\sqrt{\cot ^2(\theta ) \csc ^2(\theta )+\frac{1}{4} \left(Y+e^{-\gamma } \left({\phi_e}^2-\frac{q_m^2}{{r_h}^2}\right)-2 {\phi_e}^2+1\right)^2}}.
\end{eqnarray}
To obtain the analytic expression for $\tau$  by setting $\phi^{r_h}=0$ equivalent to zero point as
\begin{eqnarray}\label{34}
\tau= \frac{4 \pi  e^{\gamma } p^2 {r_h}^3}{\left(\pi  \lambda  {r_h}^2+1\right) \left(p^2 \left({r_h}^2 \left(-2 e^{\gamma } {\phi_e}^2+e^{\gamma }+{\phi_e}^2\right)-q_m^2\right)+3 e^{\gamma } {r_h}^4\right)}.
\end{eqnarray}
\begin{figure}
\includegraphics[width=18pc]{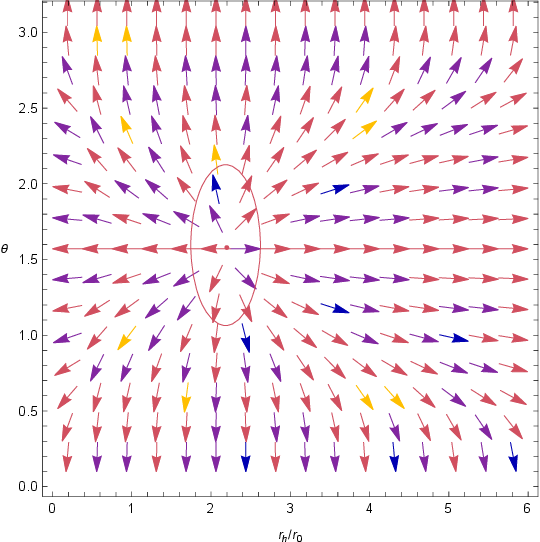}
\caption{\label{f8} Plot of unit vector field $n$ in $r_h/r_0$ vs $\theta$ plane for BH thermodynamic defects in mixed ensemble with fixed values of  $\gamma =2.3$, $\lambda =0.5$,  $p=3.2$, $q_e=1$, $q_m=1$, $\tau =30$, $a=0.13$, $b=0.78$, $r0=0.3834$ and $\phi_e=1$.  The dot represents the critical point.}
\end{figure}

In {Fig. \ref{f8}}, we plot the normalized vector in a mixed ensemble and identify a critical point. We consider the parameters $\gamma = 2.3$, $\lambda = 0.5$, and $p = 3.2$. The critical point for ${\tau}/{r_{0}} = 30$, denoted as $CP_{10}$, is located at $(a, b, r_{0}) = (0.13, 0.78, 2.3)$. The outgoing flow typically represents the small  BH phase.

\section{BH in grand canonical  ensemble}\label{sec8}

In the grand canonical ensemble, both the electric potential $\phi_e$ and the magnetic potential $\phi_m$ are kept fixed as $\phi_e=\frac{q_{e}}{r_h}$ and $\phi_m=\frac{q_m}{r_h}$, respectively. The mass parameter can be represented as
\begin{eqnarray}\label{35}
m=\frac{1}{2} \left(\frac{{r_h}^3}{p^2}+e^{-\gamma } {r_h} \left({\phi_e}^2+{\phi_m}^2\right)+{r_h}\right).
\end{eqnarray}
The modified temperature relation reads as
\begin{eqnarray}\label{36}
T= \frac{e^{-\gamma } \left(\pi  \lambda  {r_h}^2+1\right) \left(3 e^{\gamma } {r_h}^2-p^2 \left({\phi_m}^2-\left(e^{\gamma }-{\phi_e}^2\right)\right)\right)}{2 \pi  p^2 {r_h}^2}.
\end{eqnarray}
{The thermodynamic function $\phi=\frac{1}{\sin \theta} T\left(S, x^i\right)$ is computed as}
\begin{eqnarray}\label{37}
\phi=\frac{e^{-\gamma } \left(\pi  \lambda  {r_h}^2+1\right) \left(3 e^{\gamma } {r_h}^2-p^2 \left({\phi_m}^2-\left(e^{\gamma }-{\phi_e}^2\right)\right)\right)}{2 \pi  p^2 {r_h}^2 \sin (\theta )}.
\end{eqnarray}
The components of the vector field $\boldsymbol{\phi} = \left(\phi^{r_h}, \phi^\theta\right)$ are
\begin{eqnarray}\label{38}
\phi^{r_h}=\frac{e^{-\gamma } \left(\pi  \lambda  {r_h}^2+1\right) \left(3 e^{\gamma } {r_h}^2-p^2 \left({\phi_m}^2-\left(e^{\gamma }-{\phi_e}^2\right)\right)\right)}{2 \pi  p^2 {r_h}^2 \sin (\theta )},
\end{eqnarray}
and
\begin{eqnarray}\label{39}
\phi^{\theta}=-\frac{e^{-\gamma } \cot (\theta ) \csc (\theta ) \left(\pi  \lambda  {r_h}^2+1\right) \left(e^{\gamma } \left(p^2+3 {r_h}^2\right)-p^2 \left({\phi_e}^2+{\phi_m}^2\right)\right)}{2 \pi  p^2 {r_h}^2}.
\end{eqnarray}
We follow a similar procedure by plotting the unit vectors and examining the zero points by setting $\Theta = \pi / 2$ in $n^1$ and equating it to zero. The pressure value is taken below the critical pressure $P_c$ and remains unchanged. The vector $\phi$ is normalized, and its components are
\begin{eqnarray}\label{40}
\frac{\phi^{r_h}}{\|\phi\|}&=& \frac{2 \left(p^2 \left({\phi_e}^2+{\phi_m}^2\right)-e^{\gamma } \left(p^2-3 \pi  \lambda  {r_h}^4\right)\right)}
{\sqrt{4 D^2+{r_h}^2 \cot ^2(\theta ) \left(\pi  \lambda  {r_h}^2+1\right)^2 \left(e^{\gamma } \left(p^2+3 {r_h}^2\right)-p^2 \left({\phi_e}^2+{\phi_m}^2\right)\right)^2}},\\
\label{41}
\frac{\phi^\theta}{\|\phi\|}&=&-\frac{\cot (\theta ) \left(\pi  \lambda  {r_h}^2+1\right) \left(e^{\gamma } \left(p^2+3 {r_h}^2\right)-p^2 \left({\phi_e}^2+{\phi_m}^2\right)\right)}{\sqrt{4 D^2+{r_h}^2 \cot ^2(\theta ) \left(\pi  \lambda  {r_h}^2+1\right)^2 \left(e^{\gamma } \left(p^2+3 {r_h}^2\right)-p^2 \left({\phi_e}^2+{\phi_m}^2\right)\right)^2}}.
\end{eqnarray}
where $D$ is defined in the Appendix.
\begin{figure}
\begin{minipage}{18pc}
\includegraphics[width=18pc]{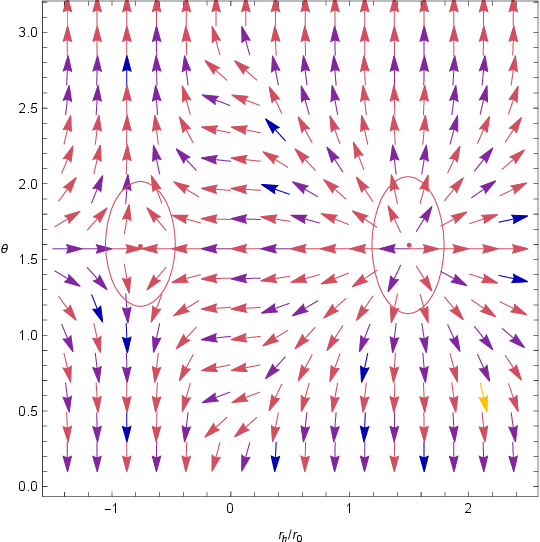}
\caption{\label{f10} Plot of unit vector field $n$ in $r_h/r_0$ vs $\theta$ plane for BH thermodynamic in grand canonical ensemble with fixed values of
$\lambda =.3$, $\gamma =1.3$, $\lambda =0.1$,  $p=2.2$, $q_e=1$, $q_m=1$, $\tau =5$, $a=0.13$ and $b=0.78$. }
\end{minipage}\hspace{3pc}%
\begin{minipage}{18pc}
\includegraphics[width=18pc]{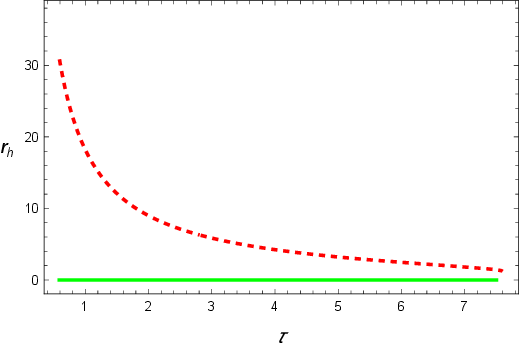}
\caption{\label{f11} Plot of  $r_h/r_0$ vs $\theta$ plane for BH thermodynamic in grand canonical ensemble with  $\gamma =1.3$, $\lambda =0.1$,  $p=2.02$, $q_e=1$ and $q_m=1$.}
\end{minipage}\hspace{3pc}%
\end{figure}
{In {Fig. \ref{f10}}, we present the normalized vector and identify two critical points. In this case, we fix $\gamma = 1.3$, $\lambda = 0.1$, and $p = 2.2$. For $\tau = 5$, there exist two critical points, denoted as $CP_{11}$ and $CP_{12}$, located at $(a, b, r_{0}) = (0.13, 0.78, -0.5)$ and $(0.13, 0.78, 1.5)$, respectively. Here, $CP_{11}$ represents the incoming flow and typically provides information about the stable phase.
}
In {Fig.~\ref{f11}}, we plot $r_h/r_0$ versus the $\theta$ plane for BH thermodynamics in the grand canonical ensemble with fixed values of $\gamma = 1.3$, $\lambda = 0.1$, $p = 2.02$, $q_e = 1$, and $q_m = 1$. When $P < P_c$, there are three on-shell BH branches, consisting of two stable BHs and one unstable BH in a specific region of $\tau$. However, when $P > P_c$, only one stable BH exists for a given $\tau$.

\section{Topological thermodynamic defects in grand canonical ensemble}\label{sec9}

\begin{figure}
\begin{minipage}{18pc}
\includegraphics[width=18pc]{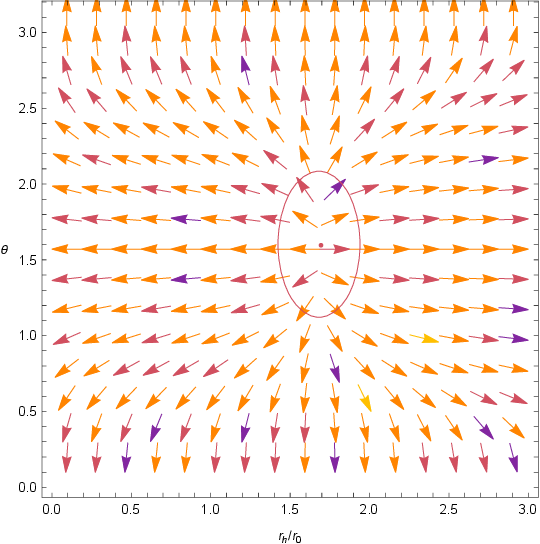}
\caption{\label{f12} Plot of unit vector field $n$ in $r_h/r_0$ vs $\theta$ plane for BH in thermodynamic defects in grand canonical ensemble with fixed values of $\gamma =3.3$, $\lambda =0.3$,  $p=1.2$, $q_e=1$, $q_m=1$ and $\tau =2$. The dot represents the critical point.}
\end{minipage}\hspace{3pc}%
\begin{minipage}{18pc}
\includegraphics[width=18pc]{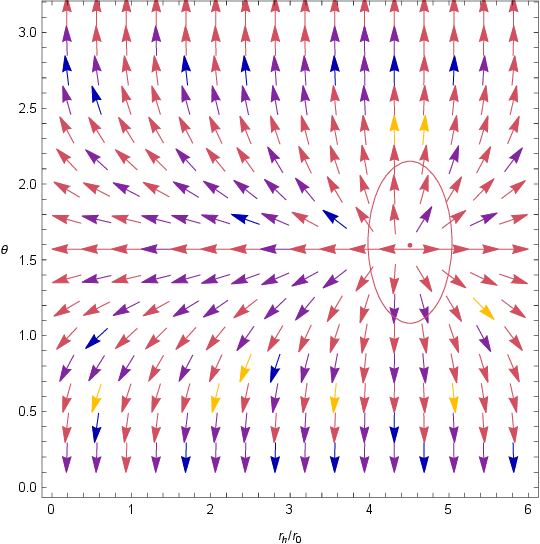}
\caption{\label{f13} Plot of unit vector field $n$ in $r_h/r_0$ vs $\theta$ plane for BH in thermodynamic defects in grand canonical ensemble with fixed values of $\gamma =1.3$, $\lambda =0.5$,  $p=4.5$, $q_e=1$, $q_m=1$ and $\tau =2$. The dot represents the critical point.}
\end{minipage}\hspace{3pc}%
\end{figure}

{To identify the  BH  as a topological defect in thermodynamics, one can utilize the following general free energy potential:
}
\begin{eqnarray}\label{42}
F=m-\frac{S}{\tau}-q_{e} {\phi_e}-q_m {\phi_m}.
\end{eqnarray}
It can be derived as
\begin{eqnarray}\label{42a}
F=\frac{1}{2} \left(\frac{r_h^3}{p^2}-\frac{2 \log \left(\pi  \lambda  {r_h}^2+1\right)}{\lambda  \tau }+e^{-\gamma } {r_h} \left({\phi_e}^2+{\phi_m}^2\right)-2 {r_h} {\phi_e}^2-2 {r_h} {\phi_m}^2+{r_h}\right).
\end{eqnarray}
From (\ref{i8}), the vector components are given by
\begin{eqnarray}\label{43}
\phi^{r_h}&=&\frac{1}{2} \left(e^{-\gamma } \left({\phi_e}^2+{\phi_m}^2\right)+Y-2 {\phi_e}^2-2 {\phi_m}^2+1\right),\\
\label{44}
\phi^{\theta}&=&-\cot (\theta ) \csc (\theta ).
\end{eqnarray} 
Two branches of critical points lead to the distinction of the topological charges when $\tau \approx 15.70$. Thus, this spacetime transforms from a  BH to a naked singularity at the topological phase. To acquire the equivalent unit vectors, one can examine them as
\begin{eqnarray}\label{45}
n^{1}=  \frac{e^{-\gamma } \left({\phi_e}^2+{\phi_m}^2\right)+Y-2 {\phi_e}^2-2 {\phi_m}^2+1}{2 \sqrt{\cot ^2(\theta ) \csc ^2(\theta )+\frac{1}{4} \left(e^{-\gamma } \left({\phi_e}^2+{\phi_m}^2\right)+Y-2 {\phi_e}^2-2 {\phi_m}^2+1\right)^2}},
\end{eqnarray}
and
\begin{eqnarray}\label{46}
n^{2}= -\frac{\cot (\theta ) \csc (\theta )}{\sqrt{\cot ^2(\theta ) \csc ^2(\theta )+\frac{1}{4} \left(e^{-\gamma } \left({\phi_e}^2+{\phi_m}^2\right)+Y-2 {\phi_e}^2-2 {\phi_m}^2+1\right)^2}}.
\end{eqnarray}
{In {Fig.~\ref{f12}}, we show that the behavior of $r_h-\tau$ for a charged modified BH in the AdS regime is separated into two entities. These parameterized forms have the same expression as stated in Eq.~(\ref{17}). For fixed values of $\gamma = 3.3$, $\lambda = 0.3$, $p = 1.2$, and ${\tau}/{r_{0}} = 2$, we obtained the critical points: $CP_{13}$ at $(a, b, r_{0}) = (0.13, 0.80, 1.7)$ and $CP_{14}$ at $(a, b, r_{0}) = (0.13, 0.80, 4.6)$, as shown in {Fig.~\ref{f13}}. One can clearly observe the regions of sign differences in the $r_h - \tau$ plot and determine the BH relation near the event horizon.
} We display the on-shell solution curve on the $r_h - \tau$ plane. By setting $\phi^{r_h} = 0$, we obtain an analytic expression for $\tau$ at the zero point as
\begin{eqnarray}\label{47}
\tau=\frac{4 \pi  e^{\gamma } p^2 {r_h}}{\left(\pi  \lambda  {r_h}^2+1\right) \left(e^{\gamma } \left(p^2 \left(-2 {\phi_e}^2-2 {\phi_m}^2+1\right)+3 {r_h}^2\right)+p^2 \left({\phi_e}^2+{\phi_m}^2\right)\right)}.
\end{eqnarray}
\begin{figure}
\begin{minipage}{18pc}
\includegraphics[width=18pc]{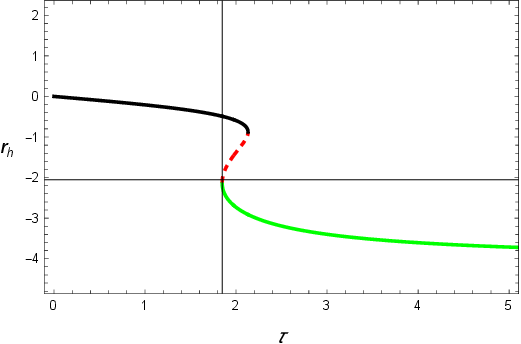}
\caption{\label{f14} Plot of  $r_h$ vs $\theta$ plane for BH thermodynamic defects in grand canonical ensemble with fixed values of $\gamma =1.3$, $\lambda =0.5$,  $p=4.5$, $q_e=1$ and $q_m=1$.}
\end{minipage}\hspace{3pc}%
\begin{minipage}{18pc}
\includegraphics[width=18pc]{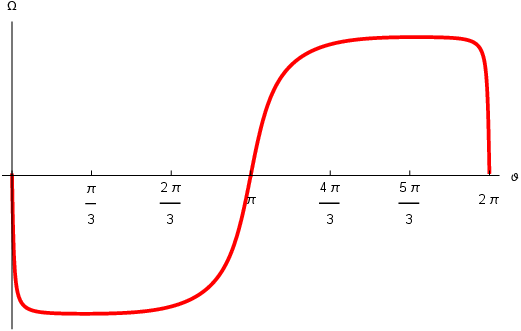}
\caption{\label{f14a} Plot of $\Omega$ vs $\vartheta$ for the contours $C_7$ with fixed values of  $\gamma =1.3$, $\lambda =0.5$,  $p=4.5$, $q_e=1$, $q_m=1$, $\tau =5$, $a=0.13$, $b=0.80$, $r_0=4.3834$ and $\phi_e=1$.}
\end{minipage}\hspace{3pc}%
\end{figure}
In {Fig. \ref{f14}}, we plot $r_h$ versus $\theta$ for BH thermodynamic defects in the grand canonical ensemble with $\gamma =1.3$, $\lambda =0.5$, $p=4.5$, $q_e=1$, and $q_m=1$. 
{In the grand canonical ensemble, for pressures below the critical pressure, this region exhibits three branches of the $\tau$ curve. The first and third branches correspond to the large and small BH phases, with positive and negative specific heat capacities, respectively.} The other branch represents the intermediate BH phase, which has a negative specific heat capacity. In this case, we observe the occurrence of a phase transition in the region where $P < P_{c}$.

Furthermore, we obtain the relationship of $\Omega$ as a function of $\vartheta$ and adopt a similar approach. In {Fig. \ref{f14a}}, we illustrate a contour $C_{7}$ with fixed values of $\gamma =1.3$, $\lambda =0.5$, $p=4.5$, $q_e=1$, $q_m=1$, $\tau =5$, $a=0.13$, $b=0.80$, $r_0=4.3834$, and $\phi_e=1$. After completing a full loop, the function $\Omega(\vartheta)$ reaches zero at $\vartheta = 2\pi$.
 {Hence, the topological charge along the contour $C_{7}$ is $0$. This indicates that the contour has no conventional critical points. Consequently, we can conclude that the naked singularity has a predicted value of $Q = 0$. Compared with Ref.~\cite{62}, this scenario is analogous to a rotating boson star devoid of an event horizon. Consequently, both are classified into the same topological class.
}

\section{Concluding remarks}\label{sec10}

{This paper presents a charged modified  BH in the AdS regime and investigates its topological thermodynamic properties using the Rényi entropy formalism. Focusing on the topological charges of BH solutions in modified gravity theories, this study aims to examine the relationship between the topology of the event horizon and the corresponding topological classes.
}

This study explores the topological thermodynamics of  BHs across canonical, mixed, and grand canonical ensembles, defining electric and magnetic potentials to capture distinct thermodynamic behaviors. {Thus, two critical behaviors have been identified in these ensembles: single generation and annihilation points, which are associated with critical pressures.}
Analyzing topological charges at these critical points reveals a conventional critical point with a topological charge of $-1$ in both the canonical and mixed ensembles. In contrast, the grand canonical ensemble exhibits a single variable topological charge of either zero or one. These findings have been compared with recent studies in Refs. \cite{27,28,30,62}, showing consistent patterns across the literature. 
The BH in the AdS regime has been identified as a topological defect within thermodynamic space, and its local and global topologies have been examined through winding numbers. Interestingly, the thermodynamic topology of the grand canonical ensemble differs from that of the canonical and mixed ensembles, suggesting unique behavior depending on the choice of ensemble.

The significance of this work extends to broader implications: it reinforces that BH solutions across various gravitational theories—whether in general relativity, modified gravity, or higher-dimensional frameworks—typically fall into one of three distinct topological classes. This categorization suggests that the presence and nature of zero points, as well as the topological number, may depend significantly on the dimensionality of spacetime. 
{Extending this approach to lower dimensions could uncover universal phenomena, which would be crucial for further exploration. Given that a positive cosmological constant is commonly accepted as a driver of cosmic expansion, these classifications highlight that BH topology depends on the choice of ensemble.}
This study aligns with existing research and provides valuable insights into BH thermodynamics within the framework of modified gravity and Rényi entropy, offering a strong foundation for future investigations in the field. 
{Moreover, this work establishes the classical topological classification of BHs across different ensembles, and it would be valuable to explore how quantum entanglement influences the topological structure of BHs in the context of modified gravity theories. This could involve examining whether the topological classification remains valid under quantum corrections and how Rényi entropy modifications interact with quantum effects near critical points.}

Extending this framework to incorporate holographic considerations may reveal deeper connections between bulk topology and boundary physics, potentially offering new insights into the AdS/CFT correspondence for modified gravity theories. Such investigations could help bridge the gap between classical and quantum descriptions of BH thermodynamics while uncovering new topological invariants that emerge solely at the quantum level.

\section*{Acknowledgement}
This project was supported by the Natural Sciences Foundation of China (Grant No. 11975145). The authors thank the reviewers for their comments on this paper.
\section*{Declaration of competing interest}
The authors declare that they have no known competing financial interests or personal relationships that could have appeared to influence the work reported in this paper.

\section*{Data Availability Statement}
This manuscript has no associated data, or the data will not be deposited. (There is no observational data related to this article. The
necessary calculations and graphic discussion can be made available on request.)

\section{Appendix}
Abbreviations of mathematical expressions are defined as follows:

\begin{eqnarray}\label{24A}
C=4 \left(e^{-\gamma } p^2 \left(q_m^2 \left(\pi  \lambda  {r_h}^2+2\right)+{r_h}^2 {\phi_e}^2\right)-p^2 {r_h}^2+3 \pi  \lambda  {r_h}^6\right)^2.
\end{eqnarray}
\begin{eqnarray}\label{40A}
D=e^{\gamma } \left(p^2-3 \pi  \lambda  {r_h}^4\right)-p^2 \left({\phi_e}^2+{\phi_m}^2\right).
\end{eqnarray}
\begin{eqnarray}\label{43A}
Y=\frac{3 {r_h}^2}{p^2}-\frac{4 \pi  {r_h}}{\pi  \lambda  {r_h}^2 \tau +\tau }.
\end{eqnarray}

\vspace{2cm}


\end{document}